\documentclass[
twocolumn,
 amsmath,amssymb,
 aps,
  twoside,
 superscriptaddress,
pra,
longbibliography
]{revtex4-1}

\usepackage{physics}
\usepackage{dsfont} 
\usepackage{color,,newfloat}
\usepackage{graphicx}
\usepackage{dcolumn}
\usepackage{bm}
\usepackage{hyperref}

\hypersetup{
    colorlinks=true,       
    linkcolor=blue,          
    citecolor=blue,        
    urlcolor=blue           
}

\usepackage{algcompatible}

\usepackage{etoolbox}

\AtBeginEnvironment{algorithm}{\vskip+15pt\noindent\par\nobreak\vskip-0pt}
\AtEndEnvironment{algorithm}{\noindent\hrulefill\par\nobreak\vskip+5pt}

\usepackage{newfloat}

\DeclareFloatingEnvironment[
    fileext=loa,
    listname=List of Algorithms,
    name=Algorithm,
    placement=tbhp,
]{algorithm}

\begin{document}


 \title{Experimental Bayesian Quantum Phase Estimation on a Silicon Photonic Chip}

\author{S. Paesani}
  \altaffiliation{These authors contributed equally to this work.}
 
\author{A.A. Gentile}%
 \altaffiliation{These authors contributed equally to this work.}
 
\author{R. Santagati}
\author{J. Wang}
\affiliation{
Quantum Engineering Technology Labs, H. H. Wills Physics Laboratory and Department of Electrical and Electronic Engineering, University of Bristol, BS8 1FD, UK. 
}

\author{N. Wiebe}
\altaffiliation{nawiebe@microsoft.com}
\affiliation{
Quantum Architectures and Computation Group, Microsoft Research, Redmond, Washington 98052, USA  
}

\author{D.P. Tew}
\affiliation{
School of Chemistry, University of Bristol, Bristol BSB 1TS, UK
}

\author{J.L. O\textquoteright Brien}
\author{M.G. Thompson}
\altaffiliation{mark.thompson@bristol.ac.uk}
\affiliation{
Quantum Engineering Technology Labs, H. H. Wills Physics Laboratory and Department of Electrical and Electronic Engineering, University of Bristol, BS8 1FD, UK. 
}

\date{\today}

\begin{abstract}
Quantum phase estimation is a fundamental subroutine in many quantum algorithms, including Shor's factorization algorithm and quantum simulation. However, so far results have cast doubt on its practicability for  near-term, non-fault tolerant, quantum devices.
Here we report experimental results demonstrating that this intuition need not be true. We implement a recently proposed adaptive Bayesian approach to quantum phase estimation and use it to simulate molecular energies on a Silicon quantum photonic device. The approach is verified to be well suited for pre-threshold quantum processors by investigating its superior robustness to noise and decoherence compared to the iterative phase estimation algorithm. This shows a promising route to unlock the power of quantum phase estimation much sooner than previously believed.

\end{abstract}

\maketitle

\textit{Introduction}---Quantum algorithms promise exponential advantages over their classical counterparts, allowing the possibility to accomplish tasks otherwise unachievable on a classical computer~\cite{Shor1994,Lloyd1996}. 
A fundamental tool in quantum computing  is the quantum phase estimation algorithm (PEA),
necessary for harnessing many of its main applications, e.g factorization of large numbers \cite{NielsenChuang,Shor1994, Lanyon2007,MartinLopez2012,Monz2016} and simulation of molecular properties \cite{ Lanyon2010,Du2010,OMalley2016,Santagati2016,AspuruGuzik2005}. 
An efficient implementation of PEA will thus be a crucial subroutine for quantum computers. 
Kitaev's iterative phase estimation algorithm (IPEA) \cite{Kitaev1996} and its adaptive 
version \cite{Griffiths1996,Dobsicek2007}
have been employed in
proof-of-principle implementations of PEA, as they solely rely on a relatively small number of qubits and logic gates \cite{Lanyon2007,Lanyon2010,Du2010,MartinLopez2012,Monz2016,OMalley2016,Santagati2016}.   
However, they require exponentially long coherence of the quantum hardware and are very susceptible to experimental noise~\cite{Dobsicek2007, Rubin2007,Wiebe2016, OMalley2016}. 
This means that conventional quantum phase estimation algorithms rapidly become  impractical if the quantum computer is not fully error-corrected, limiting the feasibility in near-term,  pre-fault-tolerant quantum machines.

A new efficient Bayesian phase estimation algorithm, called \textit{rejection filtering phase estimation} (RFPE), has been recently proposed to overcome this issue \cite{Wiebe2016}. 
The algorithm applies an approximate form of Bayesian inference to efficiently estimate the correct eigenphase.  Theoretical results suggest that RFPE has a number of desirable features: high robustness to noise, a well-motivated confidence interval for the estimated eigenphase, minuscule memory requirement for the classical control, and speed-up over the standard IPEA.

The potential of RFPE to exhibit these properties has so far been suggested by numerical simulations of the algorithm~\cite{Wiebe2016}. Experimental evidence with realistic noise sources is now required to demonstrate the high performance of the approach, potentially vindicating RFPE as a scalable, practical and quadratically faster alternative to other statistical methods~\cite{Peruzzo2014,McClean2016}.

Here we exploit a fully-reconfigurable silicon quantum photonic device, able to generate single photons on-chip and manipulate the states via an arbitrary controlled-unitary operation, to investigate the experimental viability of RFPE. The high level of precision and reconfigurability offered by the photonic chip allow us to compare and contrast the performance of both RFPE and IPEA under the action of different controllable experimental noises. 
Specifically, we find experimental evidence that RFPE is robust against realistic sources of errors, making it very appealing for near-term useful applications.

\begin{figure*}[th!]
  \centering
  \includegraphics[width=0.85 \textwidth]{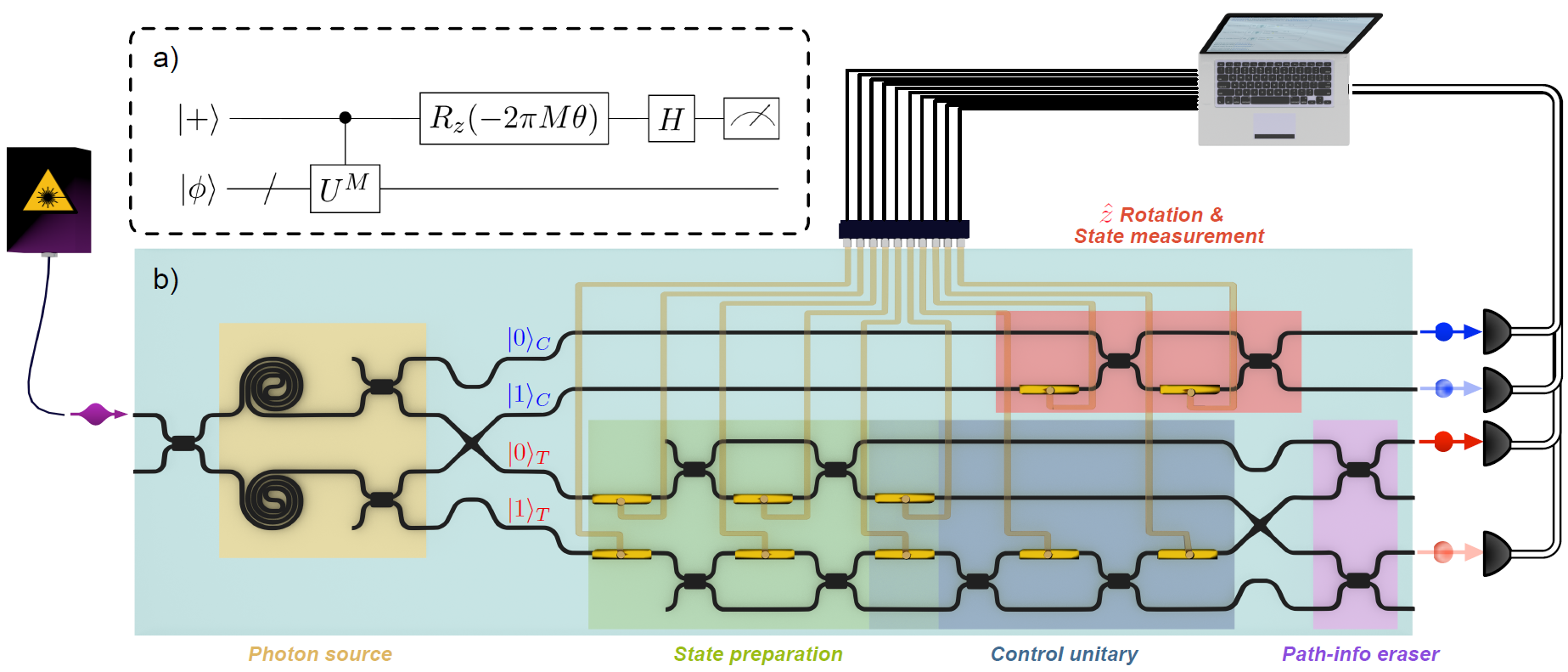}
  \caption{
  a). Quantum circuit for standard iterative and Bayesian phase estimation algorithms. 
  b). Experimental setup and the integrated silicon quantum photonic device. The quantum chip can perform any controlled-$U(2)$ operation and any single-qubit state preparation and analysis. Photons are produced and guided in the silicon waveguides (black wires) and reconfigurably controlled by thermo-optical phase shifters. Coherent light was used to generate photons and superconducting nanowire detectors were used for the detection, both coupled to the chip through lensed-single mode fibers. The implementation of the algorithms was achieved by interfacing the quantum device with a classical CPU. } 
  \label{Fig:chip}
\end{figure*}

\textit{Phase estimation}---
The goal of phase estimation is, given 
a unitary $\hat U$ and a quantum state $\ket{\psi}$, to learn an eigenvalue $e^{i\phi}$ of $\hat U$ within the support of $\ket{\psi}$. Standard algorithms work by interfering paths in which either $\hat{\openone}$ or $\hat U^j$ is applied to $\ket{\psi}$, for integer values of $j$, and then recombining the paths together to allow them to interfere~\cite{NielsenChuang}.  Iterative phase estimation works by pooling the results of many such experiments and using a classical inference algorithm to estimate an eigenphase of $\hat U$ from the resultant interference pattern.

IPEA allows eigenvalues of $\hat U$ to be learned quadratically faster than by statistical sampling and requires exponentially fewer measurements.
However, it typically requires long evolution times which can reduce its utility in pre--fault tolerant hardware~\cite{Dobsicek2007,Griffiths1996,Kitaev1996}. The most commonly used IPEA algorithm works by inferring each of the bits in a binary expansion of the eigenphase $\phi$ in reverse order~\cite{Kitaev1996,Griffiths1996,Dobsicek2007}. The method uses the circuit in Figure~\hyperref[Fig:chip]{\ref*{Fig:chip}a}, where the measurement on the control qubit gives output $0$ or $1$ with probabilities $\cos^2(\pi M [\phi-\theta])$ and $\sin^2(\pi M [\phi-\theta])$, respectively. As each bit is learned iteratively, the algorithm applies a fixed policy for updating $M$ and $\theta$ (see Appendix~\hyperref[pseudocodeIPEA]{\ref*{pseudocodeIPEA}}). 

RFPE is in many ways simpler. This Bayesian approach uses a Gaussian probability distribution $P(\phi)$ (the prior) representing the confidence that the current hypotheses is the correct eigenphase. The result of each new measurement is used to update the
mean $\mu$ and standard deviation $\sigma$ according to Bayes' theorem, approximated with rejection sampling (see Appendix~\hyperref[pseudocode]{\ref*{pseudocode}}). Specifically, a host of particles are drawn from the prior distribution and then probabilistically discarded  based on the likelihood function.  The remaining samples model the posterior probability distribution, which becomes the new prior.  Since the number of surviving particles decreases exponentially, the posterior distribution is refit to a new Gaussian at each step and
fresh particles added, drawn from this new distribution. 
Rather than learning each of the bits of $\phi$ individually, the RFPE algorithm gains information about every bit simultaneously.

In more detail, if an outcome  $E\in\{0,1\}$ is obtained from an experiment with parameters $M$ and $\theta$, the likelihood function  for the host of particles $\{x_i\}$ is calculated:
$$P(E|x_i;M,\theta)/\kappa=\begin{cases}\cos^2(M\pi(x_i - \theta)) /\kappa& E=0\\
\sin^2(M \pi (x_i - \theta))/\kappa& E=1, \end{cases}$$ where $\kappa$ is a rescaling constant \cite{Wiebe2016}.
Particles from the prior are discarded probabilistically based on this function.
In order to maximize the information gain at each step, new values of $M$ and $\theta$ can be extracted from the prior distribution using various optimization methods. A near-optimal choice 
is provided by particle guess heuristics \cite{Ferrie2013,wiebe2014hamiltonian}, giving $\theta\sim P(\phi)$ and $M=\lceil 1.25/ \sigma \rceil $.  
If the likelihood function fails to describe the data (due to experimental noise) then the algorithm estimates the best model for the experimental results within the assumptions, and gives a firm estimate of the uncertainty in the eigenphase. 
While in standard IPEA any error that occurs in inferring a bit cannot be corrected in subsequent algorithm steps,  RFPE does not suffer this issue because it does not try to infer the bits sequentially.
 While these expectations have been born out in simulation~\cite{Wiebe2016}, they have not been verified in practice.  We provide such verification below using an integrated quantum photonic device. Further details for RFPE and IPEA can be found in Appendix~\hyperref[pseudocodeIPEA]{\ref*{pseudocodeIPEA}} and  Appendix~\hyperref[pseudocode]{\ref*{pseudocode}}.

\textit{Integrated quantum photonic device} --- Silicon quantum photonics have emerged as a promising approach for the realisation of quantum hardware, since in principle all the necessary photonic components (sources, circuits, filters and detectors) for quantum information processing can be integrated on a single platform~\cite{silverstone2016}.  
We developed a quantum photonic device in silicon waveguides for the experimental implementation of RFPE, shown in Fig.~\hyperref[Fig:chip]{\ref*{Fig:chip}b}. The chip was manufactured on a standard Silicon-on-Insulator platform and is capable of performing arbitrary two-qubit controlled unitary operations.
Using spontaneous four-wave mixing,
photon pairs were created in two spiral sources pumped with $\simeq 20$~mW bright light near $1550$~nm wavelength~\cite{Reid1986}. 
Separating the photons by multi-mode interferometer beam-splitters (BS) and swapping the two modes by a waveguide crosser yield a maximally entangled two-photon post-selected state $(\ket{0}_C\ket{0}_T + \ket{1}_C\ket{1}_T) / \sqrt{2}$, where $\ket{0}$ ($\ket{1}$) indicates the photon state either in its upper or lower spatial mode~\cite{Wang2015, Silverstone2015}, whereas $C\
(T)$ subscript refers to the control (target) register.
Two additional spatial modes are then added to the target register, obtaining a path encoded qubit for each of the two modes of the target wave-function. After an initial state preparation $|\phi\rangle_T$, each qubit is 
manipulated with a separate transformation, depending on which path the photon is travelling on: 
the identity $\hat{\mathds{1}}$ for the upper modes (the ones corresponding to a $\ket{0}_C$) and an arbitrary unitary $\hat{V}$ for the bottom modes (the ones corresponding to a $\ket{1}_C$). 
The path information is erased by two BSs and the state is finally post-selected obtaining the equivalent photon count statistics of an arbitrary control-unitary operation $(\ket{0}_C \otimes \ket{\phi}_T + \ket{1}_C \otimes \hat{V} \ket{\phi}_T)/ \sqrt{2}$ \cite{Santagati2016, Zou2013}.  The quantum logic of Fig.~\hyperref[Fig:chip]{\ref*{Fig:chip}a}  is completed by performing a single qubit operation on the control photon. 
All the processes required for state preparation, manipulation and measurement are achieved through thermo-optical phase shifters and Mach-Zehnder interferometers, as shown in Fig.~\hyperref[Fig:chip]{\ref*{Fig:chip}b} \cite{Silverstone2015,Reed2010}. Controlling the electric power supplied to the phase shifters, each phase $\varphi$ can be driven with an average precision of $\simeq 0.01 $~rad (see Appendix~\ref{Setup} and Appendix~\ref{SimErrors}).
Finally, photons were detected by superconducting nanowire single photon detectors (SNSPD) and coincidence counts obtained by a time interval analyser.
The photon statistics were used to measure the projectors $\Pi({0,1})$ on the computational basis of the photonic qubits. The Bayesian update and changes to the controls of the quantum device required by RFPE can be calculated and fed to the quantum system using an interfaced classical computer.

\begin{figure}
  \centering
  \includegraphics[width=0.45\textwidth]{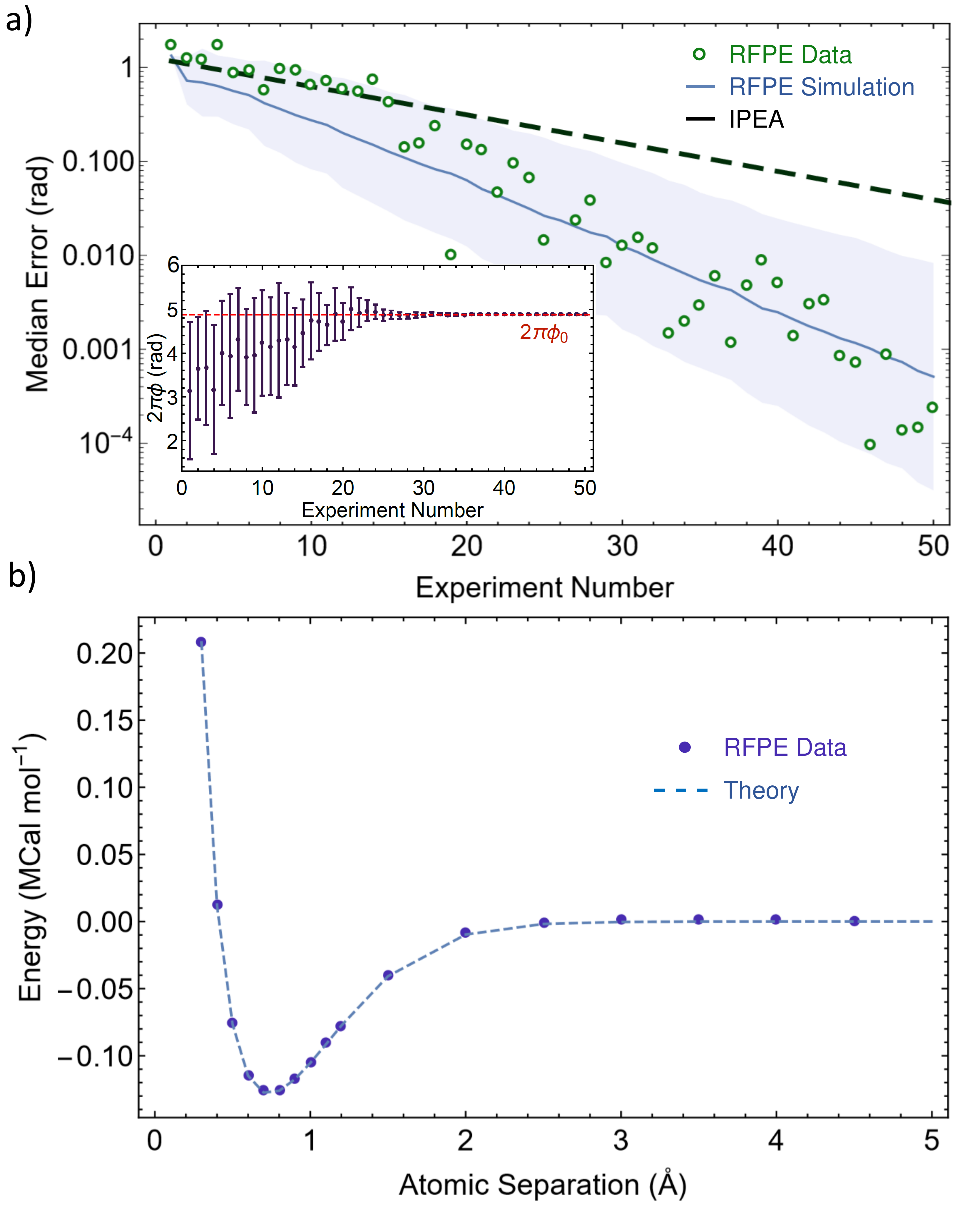}
  \caption{ a) Convergence of RFPE  to the true phase value $2\pi\phi_0=4.8741$~rad related to the energy of the dissociated $\text{H}_2$ molecule. The initial prior distribution is $\mathcal{N}(\pi,\pi^2)$.
Data points show an exponential shrink of the error, compatible with simulations (blue line) of the device performance averaged over 1000 runs of the RFPE algorithm (shaded area: 67.5\%  credible interval). The dashed black line denotes the convergence of the standard IPEA using the parameters discussed in the main text. 
Inset: convergence of the phase estimation to $\phi_0$ (red line), where errors are evaluated using the s.d. of the prior distribution.
b) Bonding energies of the $\text{H}_2$ molecule for various atomic distances using RFPE with 50 steps. Energy estimations are achieved within  chemical accuracy. Errors are smaller than the markers and neglected in the plot for more clarity. The dashed line represents theoretical energies. }\label{Fig2}

\end{figure}

\begin{figure*}[th!]
  \centering
\includegraphics[width=0.95 \textwidth]{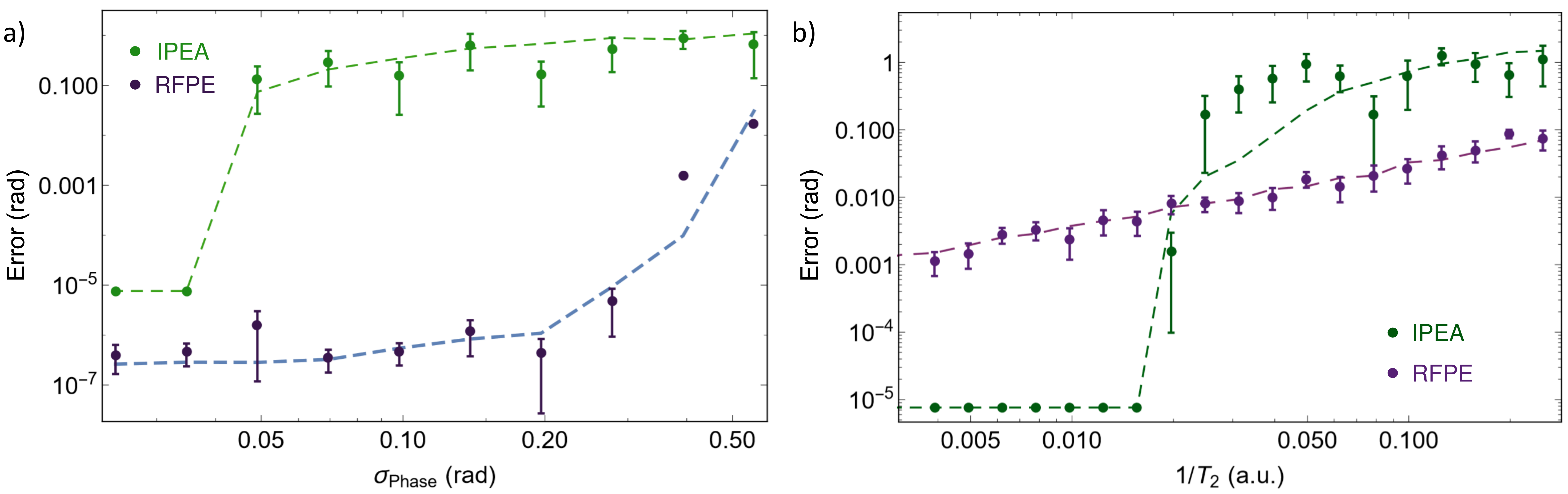}
  \caption{Effects of different experimental noises on phase estimation strategies. 
a) Infidelity of quantum operation.
Each of the correct phases $\bar{\varphi}_i$ for the phase gates is synthetically replaced with a Gaussian distributed $\varphi_i \sim \mathcal{N}(\bar{\varphi}_i,\sigma_{Phase})$,  where $\sigma_\text{phase}$ represents a noise in the phases. 
b)
Decoherence.
For IPEA data, experiments were repeated $10$ times with 16-bit accuracy to evaluate median error and error bars,
while the RFPE data were collected from a single run, after 100 measurements, and directly used to evaluate the error and uncertainty within the algorithm. 
Error bars for the estimated phase represent in both plots either a $67.5\%$ credible region for RFPE, either a $67.5\%$ confidence interval for IPEA.  In the cases where error bars are smaller than the markers they have been omitted for clarity.
Points are experimental data and dashed lines are simulations averaged over 1000 runs. 
The simulations take into account the characterized residual phase noise in the device. 
}\label{Fig3}
\end{figure*}

\begin{figure}[th!]
   \centering
 \includegraphics[width=0.45\textwidth]{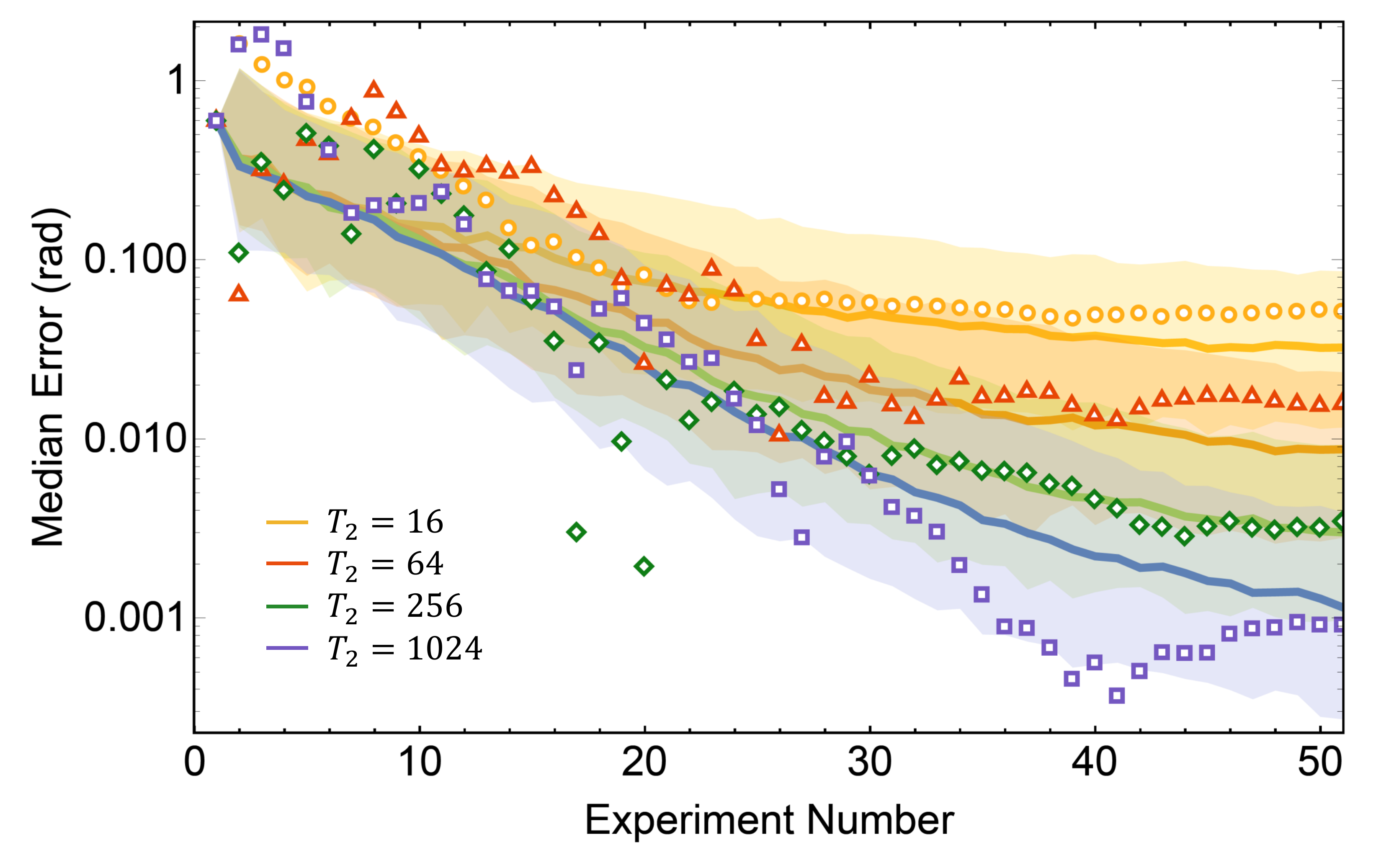}
 \caption{Experimental data (points) and simulation (lines) of the behaviour of RFPE under different decoherence times $T_2$. Shaded areas represent a 67.5\% credible interval from data simulated averaging over 500 runs of RFPE.}\label{Fig4}
 \end{figure}

\textit{Experimental results} --- The rapid reconfigurability and the high precision of the silicon photonic device were crucial to the practical implementation and testing of RFPE.  
As is usual in photonic experiments, where measurements provide probability distributions rather than single-shot data, the value of $E$ was determined by the majority voting method on the projective measurement statistics for both RFPE and IPEA  (see Appendix~\ref{MethodsDiscussion}).
For the stochastic representation of the prior, $1000$ particles was found to be reliable.
The initial prior distribution was set to a Gaussian $\mathcal{N}(\pi,\pi^2)$, which approximates a uniform prior.

In Fig.~\hyperref[Fig2]{\ref*{Fig2}a} we report the results from a single RFPE run, which demonstrates that the estimation converges exponentially to the true eigenphase. 
For this case an error as low as $2.4 \cdot 10^{-4}$~rad is achieved within 50 experiments on the quantum device. This error is in good agreement with the standard deviation of the final posterior Gaussian distribution ($\simeq 4.2 \cdot 10^{-4}$~rad after 50 experimental steps, see the inset of Fig.~\hyperref[Fig2]{\ref*{Fig2}a}), thus confirming that the algorithm provides a reliable uncertainty estimate. 
We remark that this feature is not present in other phase estimation techniques, which do not provide a rigorous estimation of the phase uncertainty that arises in non-fault-tolerant devices~\cite{Wiebe2016}. 
A possible strategy for obtaining such uncertainty estimates from IPEA would be to determine the mean and the standard deviation of the measured eigenphase from repeated experiments.
Fig.~\hyperref[Fig2]{\ref*{Fig2}a} shows that when adopting this strategy with a reasonable cardinality for the experiments\textcolor{blue}, $10$ for the curve reported, RFPE outperforms IPEA. 

To extensively test how RFPE performs in key applications, we scanned the energy of molecular $\text{H}_2$ for different atomic separations, using $50$ iterations of RFPE for each eigenphase evaluation. The eigenstates of the molecular Hamiltonian are mapped into a the qubit basis using the Jordan-Wigner transformation and the eigenphases are directly related to the binding energy \cite{AspuruGuzik2005,Lanyon2010,OMalley2016,Santagati2016}. As shown in Fig.~\hyperref[Fig2]{\ref*{Fig2}b}, the estimated energies match the theoretical values with high precision. The average error for the dataset is  $0.72$ kCal/mol, providing a precision higher than chemical accuracy ($\simeq 1$ kCal/mol).   

The main advantage of the Bayesian approach over traditional methods comes from its expected reliability on non-fault-tolerant devices. 
Here we investigate experimentally the robustness of the protocol against two main controllable sources of noises: \textit{gate errors} and  \textit{decoherence}.

The infidelity of unitary operations is a well known problem existing in quantum hardware, and is typically given by a noisy control and imperfect manufacture and calibration of the components. 
On integrated quantum photonic devices it is mainly due to control noise and residual crosstalk
of the phase gates, which 
are expected to occur on the device in Fig.~\ref{Fig:chip} \cite{Santagati2016}. 
Our electronic phase-shifters driver allows to precisely control the thermo-optical phase gates as residual thermal cross-talk can be compensated by further calibration (see Appendix~\ref{SimErrors}). 
The high controllability allows us to add a tunable level of noise on the phase gates to study the robustness of RFPE. This is achieved by replacing the correct phases $\bar{\varphi}$ required to implement the unitary transformation with synthetic values $\varphi$ sampled from a Gaussian distribution $\varphi \sim \mathcal{N}(\bar{\varphi},\sigma_\text{phase})$. 
The parameter $\sigma_\text{phase}$ mimics a Gaussian noise in the phases, which in turn introduces a controllable noise in the fidelity of both the implemented state preparation and the unitary evolution. 
Fig.~\hyperref[Fig3]{\ref*{Fig3}a} shows the convergence of both RFPE and  IPEA scanning $\sigma_\text{phase}$ up to $0.55$~rad, corresponding to average state fidelity $94\%$ and gate fidelity $91\%$ expected in the chip (see Appendix~\ref{SimErrors}). We report the performance of RFPE with 100 steps, compared to a 16-bit IPEA, averaged over 10 runs to estimate the error bars, i.e. 160 total experiments per data point. 
We remark that since a reasonable error bar estimation requires a higher number of experiments for IPEA than RFPE, the relative rates of convergence are not immediately obvious from these figures. Rather, these plots illustrate how the performance of each algorithm is affected by increasing errors, to compare their robustness to noise.
For $\sigma_\text{phase}\geq 0.05$~rad IPEA  dramatically decreases in accuracy and becomes quickly unreliable. This occurs because while the majority voting scheme provides error resilience for small error rates, it can diverge rapidly once the error rate crosses a threshold (see Appendix~\ref{IPEAerrors}).  

On the other hand, in this regime the performance of RFPE is initially only slightly affected, maintaining a very high level of precision even when IPEA fails.  This is expected because RFPE does not make hard decisions about bits as each experiment yields information about all bits at once.  Thus errors are unlikely to be critical.  In order for RFPE to be substantially affected we require $\sigma_\text{phase}$ higher than $0.3$~\ rad, a value much higher than the actual experimental noise in our device.

Decoherence is an important limitation in many quantum computing experiments but
it plays a minor role in integrated quantum photonics. 
It must then be introduced artificially in our experiment. In order to simulate it, coincidence counts provided by the SNSPDs for the $\Pi(0)$ and $\Pi(1)$ projectors were progressively flattened out by classical post-processing and combined with Poissonian noise in the measurements. In agreement with the depolarizing noise model we mimicked the presence of a normalised  decoherence time $T_2$ (see Appendix~\ref{DecTime}) such that: 
\begin{equation*}
P'(D|\phi)=e^{-M/T_2}P(D|\phi)+\frac{1-e^{-M/T_2}}{2},
\end{equation*}
where $P(D|\phi)$ is the data obtained from the photon coincidence counts for the outcome $D\in\{0,1\}$. We introduce this noise model while processing online the output data during the iterative process, thus affecting the choice of the experiments. In this way it is possible to simulate the behaviour of RFPE and IPEA in systems that are prone to this model of decoherence.  

We studied the action of the depolarizing noise up to $T_2=4$ for both IPEA and RFPE, as shown in Fig.~\hyperref[Fig3]{\ref*{Fig3}b}.
The performance of IPEA has a substantial and sharp deterioration at $T_2$ close to 32, whereas the median error of 100-step RFPE decreases only polynomially with $1/T_2$, maintaining an error $\mathcal{O}(10^{-2})$ even in the regime where conventional IPEA fails to provide any reliable estimate of the phase. 
In the presence of characterized depolarizing noise an optimized value for $M$ is given by $\text{min}(\lceil 1.25/\sigma\rceil,T_2)$~\cite{Wiebe2016}, which, however, implies that,
when decoherence is significant, the performance of RFPE degrades significantly. This behaviour is exhibited by the experimental data in Fig.~\ref{Fig4}, where the convergence of RFPE is reported under the action of various $T_2$.
We observe that RFPE ceases to learn exponentially quickly when $1/\sigma \sim T_2$, after which the algorithm continues to learn at a polynomial rate, unlike IPEA~\cite{Wiebe2016}.

\textit{Conclusion}---Our work shows how the precision and controllability developed in quantum technologies, here in particular integrated photonics, allows to go beyond the basic proof-of-principle demonstrations of quantum algorithms and to enter a regime where they can be extensively tested and compared. We experimentally verified the Bayesian phase estimation algorithm on a fully programmable silicon quantum photonic device and demonstrated its superior performance in presence of noise. Although in this work the experiment is performed using a small-scale unitary and a photonic device, more complex future implementation can be efficiently performed on any scalable quantum architecture. 
The Bayesian approach remarkably lowers the requirements for the implementation of quantum phase estimation on pre-fault-tolerant devices, showing a new promising route for practical and useful quantum information processing in the near future. 

\textit{Acknowledgements}---We thank A. Laing, J.W. Silverstone, D. Bonneau, C. E. Granade and X. Zhou for insightful discussion. 
We thank K. Ohira, N. Suzuki, H. Yoshida, N. Iizuka and M. Ezaki for the device fabrication.  
We acknowledge the support from the
Engineering and Physical Sciences Research Council (EPSRC), European Research Council (ERC), Photonic Integrated Compound Quantum Encoding (PICQUE), FP7 Action: Beyond the Barriers of Optical Integration (BBOI), Quantum Simulation on a Photonic Chip (QuChip), US Army Research Office (ARO), and Centre for Nanoscience and Quantum Information (NSQI). D.P.T thanks the Royal Society for a University Research Fellowship. J.L.O'B. acknowledges a Royal Society Wolfson Merit Award and a Royal Academy of Engineering Chair in Emerging Technologies.
M.G.T. acknowledges the support from an EPSRC Early Career Fellowship and the Toshiba Research Fellowship scheme.

\bibliography{biblio}

\onecolumngrid
\clearpage

\appendix

\onecolumngrid

\section{Iterative Phase Estimation Algorithm (IPEA)}
\label{pseudocodeIPEA}
In the standard implementation of IPEA less significant digits of the unknown phase are evaluated first using the circuit in Fig.~1A. The information obtained by the measurement is transferred to the quantum circuit and allows to refine the measurement of more significant digits via an $R_z$ rotation on an ancillary qubit. This procedure increases the probability of obtaining the correct phase value. If the phase has a binary expansion with a finite number of digits, say $n$ bits, the algorithm is deterministic and requires $n$ iterations of the experiment. 

Each digit of the eigenphase is iteratively inferred, starting from the least significant one. The exponent $M$ is chosen so that, if we want to obtain $\phi$ with $n$ bits precision, at the $j$-th iteration we have $M_j=2^{n-j}$. Supposing that $\phi$ has the finite binary expansion $\phi=0.\phi_1\ldots\phi_n$, the first iteration step is performed using $\theta_1 = 0$ and $M_1=2^{n-1}$, which from Eq.~\ref{AppendixP0} and Eq.~\ref{AppendixP1} results in $P(0)=\cos^2(\pi\  0.\phi_n)$. The data obtained from the measurement is 0 if $\phi_n=0$ and 1 if $\phi_n=1$, inferring the last bit of $\phi$ deterministically. For the $j$-th iteration we choose $\theta_j=0.0\phi_{n-j+2}\ldots\phi_{n-1}\phi_{n}$ so that, with a same reasoning as before, the digit $\phi_{n-j+1}$ is exactly determined from the projective measurement of the ancilla. Iterating this process $n$ times we can reconstruct the whole binary expansion of $\phi$. 

In presence of experimental noises, the probability of measuring the correct value of the $k$-th bit for single-shot measurements, given in Ref.\cite{Dobsicek2007}, is
\begin{equation}
P_k(\Delta_x,T_2) =\frac{ \left(   1+ e^{-\Delta_x^2 - a 2^k  T_2} \right)}{2},
\label{Dobs}
\end{equation}

where the errors considered here are a dephasing with decoherence time $T_2$ and $R_x(\delta)$ errors on the control qubit for the state preparation  and measurement projection, with the angle $\delta_x$ Gaussianly distributed with variance $\Delta_x$. $a$ is a system depended parameter \cite{Dobsicek2007}. The susceptibility of the algorithm to experimental noise  is evident from Eq.\ref{Dobs}. The success probability can be increased repeating the measurements for each qubit and adopting a majority voting scheme, as we do in our implementation, but still maintains an exponential scaling with the noise level. Moreover, if an error occurs, IPEA is not able to correct it and has limited capability to recover from it. Therefore, as noise become significant, errors cumulate and make the algorithm fail.

\section{Rejection Filtering Phase Estimation (RFPE)}
\label{pseudocode}
We describe the Bayesian phase estimation algorithm,  RFPE, introduced in \cite{Wiebe2016} and employed in our experiment. 
At a fundamental level, RFPE involves the exact same experimental setups as traditional iterative phase estimation.  The main differences arise in the way RFPE chooses experiments and also how the experimental data is processed.

The fundamental object in RFPE, as well as all other Bayesian approaches, is the prior distribution $P(\phi)$.  It describes the algorithms confidence that a particular eigenphase $\phi\in [0,2\pi)$ corresponds to the eigenvector provided to the algorithm.  As experiments are performed, the prior distribution is then updated based on the likelihood of the observed data.

Bayes' theorem provides the correct way to update such prior beliefs when provided with evidence.  The theorem states that if an event $E$ is observed for an experiment parametrized by $M$ and $\theta$ then the updated prior distribution should be
\begin{equation}
P(\phi|E;M,\theta)=\frac{P(E|\phi;M,\theta)P(\phi)}{\int P(E|\phi;M,\theta)P(\phi)d\phi},
\label{BayesUpdate}
\end{equation}
where $P(E|\phi;M,\theta)$ is known as the likelihood function and the updated prior distribution $P(\phi|E;M,\theta)$ is known as the posterior distribution.

The central challenge of Bayesian phase estimation is that even if the prior distribution is discretized then the complexity of storing the prior distribution to sufficient accuracy can be immense.  For example, if error on the order of $10^{-9}$ is needed, then several gigabytes of memory will be needed to store the distribution.  This makes it impractical for use in Shor's algorithm.  Also, if the data arises from more than one eigenvalue then it is necessary to store multiple hypotheses.  This exponentially increases the number of hypotheses that need to be tracked, rendering such Bayesian inference impractical.

RFPE avoids this problem by invoking a form of assumed density filtering that is based on rejection sampling.  This approach allows inference to be performed quickly in a memory limited environment without the need to compute costly functions.  The algorithm works by positing a model for the prior distribution, which for convenience is usually taken to be a Gaussian.  The goal of each update is then to find the best model for the posterior distribution within the family of allowed models.  The procedure involves drawing a sample from the prior distribution and then computing the likelihood that the data would emerge if that eigenphase were correct.  The sample is then randomly accepted with probability equal to this likelihood.  The mean and covariance matrix of these accepted samples can easily be shown to match the posterior mean and covariance.  Therefore these statistics provide the best model for the posterior distribution within any family of distributions that are parametrized by their first two moments.   This approach is therefore well suited for experiment because it naturally outputs an estimation for its uncertainty in the form of the covariance matrix, or variance, of the distribution.

The likelihoods used in this algorithm correspond to the measurement probabilities predicted by quantum mechanics for the circuit in Fig.~1A of the main text. Specifically, if an experiment with parameters $(M,\theta)$ is performed, the measurement on the ancillary qubit will have outcome $E\in\{ 0,1 \}$ with probability:
\begin{align}
   P(0|\phi;M,\theta)&=\cos^2(\pi M [\phi-\theta]), \label{AppendixP0}\\
   P(1|\phi;M,\theta)&=\sin^2(\pi M [\phi-\theta]).\label{AppendixP1}
\end{align}

An important remaining issue is that of choosing the parameters $M$ and $\theta$ that, given the prior distribution, optimize the information gain from the experiment. In our experiment we have used the Particle Guess Heuristic (PGH)~\cite{Ferrie2013,wiebe2014hamiltonian}, which provides near-optimal values and does not require any precomputation. In this approach the experiment is given by $M=\lceil1.25/\sigma\rceil$, $\theta\sim \mathcal{N}(\mu,\sigma)$. The fact that these parameters depend intimately on the current estimate of the uncertainty means that RFPE is implicitly an adaptive protocol.  Furthermore, the fact that RFPE uses prior information means that the asymptotic optimality result of~\cite{Ferrie2013}, which is based on the Cramer-Rao bound, does not apply in this setting.
It was further shown that this heuristic approximately saturates the Bayesian Cramer-Rao bound~\cite{Wiebe2016} under the assumption of a Gaussian prior and thereby demonstrating that the PGH is asymptotically near-optimal under these assumptions.

The full RFPE algorithm is described in Algorithm~\ref{PseudoRFPE}. For further details regarding, for example, the number of particles required to achieve small errors and variance reduction strategies, we refer to \cite{Wiebe2016}.

\begin{algorithm}
 \hrulefill
   \vskip-8pt
\caption{Bayesian update via Rejection Filtering}
  \vskip-7pt
  \hrulefill
\label{update}
\begin{algorithmic}
\STATE  \textbf{Input:} experimental data $E$, mean $\mu$ and variance $\sigma$ of the prior distribution, experiment parameters $M$ and $\theta$, number of particles $N_{\text Part}$, scale $\kappa_E$.
\STATE  \textbf{Output:} mean $\mu '$ and variance $\sigma '$ of the posterior distribution. \\

\STATE \textbf{function} $\text{UPDATE}_{\text RF}(E,\mu,\sigma,M,\theta,N_{\text Part},\kappa_E)$ 
 
 \STATE \indent $\mu_{\text acc}, \mu'_{\text acc}, V_{\text acc}, V'_{\text acc},N_{\text acc}=0.$ \indent $\triangleright$ Initialize accumulators.
 
\STATE \indent \textbf{for} $i\in 1 \rightarrow N_{\text Part}$ \textbf{do}
 
\STATE \indent \indent $x\sim \mathcal{N}(\mu,\sigma)$. \indent $\triangleright$ Sample a particle according to prior distribution.

\STATE \indent \indent $x= x \text{ mod } 2\pi$.
\STATE \indent \indent $x'= x+\pi \text{ mod } 2\pi$.

\STATE \indent \indent $u\sim \text{Uniform}(0,1)$. \indent $\triangleright$ Accept the particle with probability $P(E|x)$.

\STATE \indent \indent \textbf{if} $P(E|x)\geq \kappa_E u$ \textbf{then}

\STATE \indent \indent \indent $\mu_{\text acc}=\mu_{\text acc}+x$.
\STATE \indent \indent \indent $V_{\text acc}=V_{\text acc}+ x^2$.
\STATE \indent \indent \indent $V'_{\text acc}=V_{\text acc}+ x'^2$.
\STATE \indent \indent \indent $N_{\text acc}=N_{\text acc}+1$.

\STATE \indent \indent \textbf{end if}

\STATE \indent \textbf{end for}
 
\STATE \indent $\mu'=\mu_{\text acc}/ N_{\text acc}$.\indent $\triangleright$ Calculate mean and variance of the posterior distribution using accepted particles.
 
\STATE \indent $\sigma'=\text{min}\left( \sqrt{(V_{\text acc}- \mu_{\text acc}^2)/(N_{\text acc}-1)},\sqrt{(V'_{\text acc}- \mu_{\text acc}^2)/(N_{\text acc}-1)} \right)$.
 
 \STATE \indent \textbf{return} $(\mu ',\sigma ')$.
 \STATE \textbf{end function}
\end{algorithmic}
\end{algorithm}

\begin{algorithm}
 \hrulefill
 \vskip-8pt
\caption{Rejection Filtering Phase Estimation (RFPE) }
  \vskip-7pt
  \hrulefill
\label{PseudoRFPE}
\begin{algorithmic}
\STATE  \textbf{Input:} Initial prior distribution $\mathcal{N}(\mu_0,\sigma_0)$, total number of experiments $N_{\text Steps}$, number of particles $N_{\text Part}$, scale $\kappa_E$.
\STATE  \textbf{Output:} the phase estimation $\mu$ and its uncertainty $\sigma$.\\

\STATE $\mu=\mu_0$, $\sigma=\sigma_0$. \indent $\triangleright$ Initialize the mean $\mu$ and variance $\sigma$ of the distribution.

\FOR{$i\in 1 \rightarrow N_{\text Steps}$}
\STATE $M=\lceil 1.25/\sigma \rceil$, $\theta\sim \mathcal{N}(\mu,\sigma) $. \indent $\triangleright$ Get approximate optimal parameters $M$ and $\theta$ for the experiment via PGH.
\STATE Get data $E$ from the experiment using parameters $M$ and $\theta$.
\STATE $( \mu,\sigma )=\text{UPDATE}_{\text RF}(E,\mu,\sigma,M,\theta,N_{\text Part},\kappa_E)$. \indent $\triangleright$ Perform a Bayesian update via Rejection Filtering.  
\ENDFOR 
\STATE \textbf{return} $(\mu,\sigma)$.

\end{algorithmic}
\end{algorithm}

\section{Experimental setup}
\label{Setup}

\subsection{The silicon device and experimental details}

\begin{figure}
   \centering
 \includegraphics[width=0.8\textwidth]{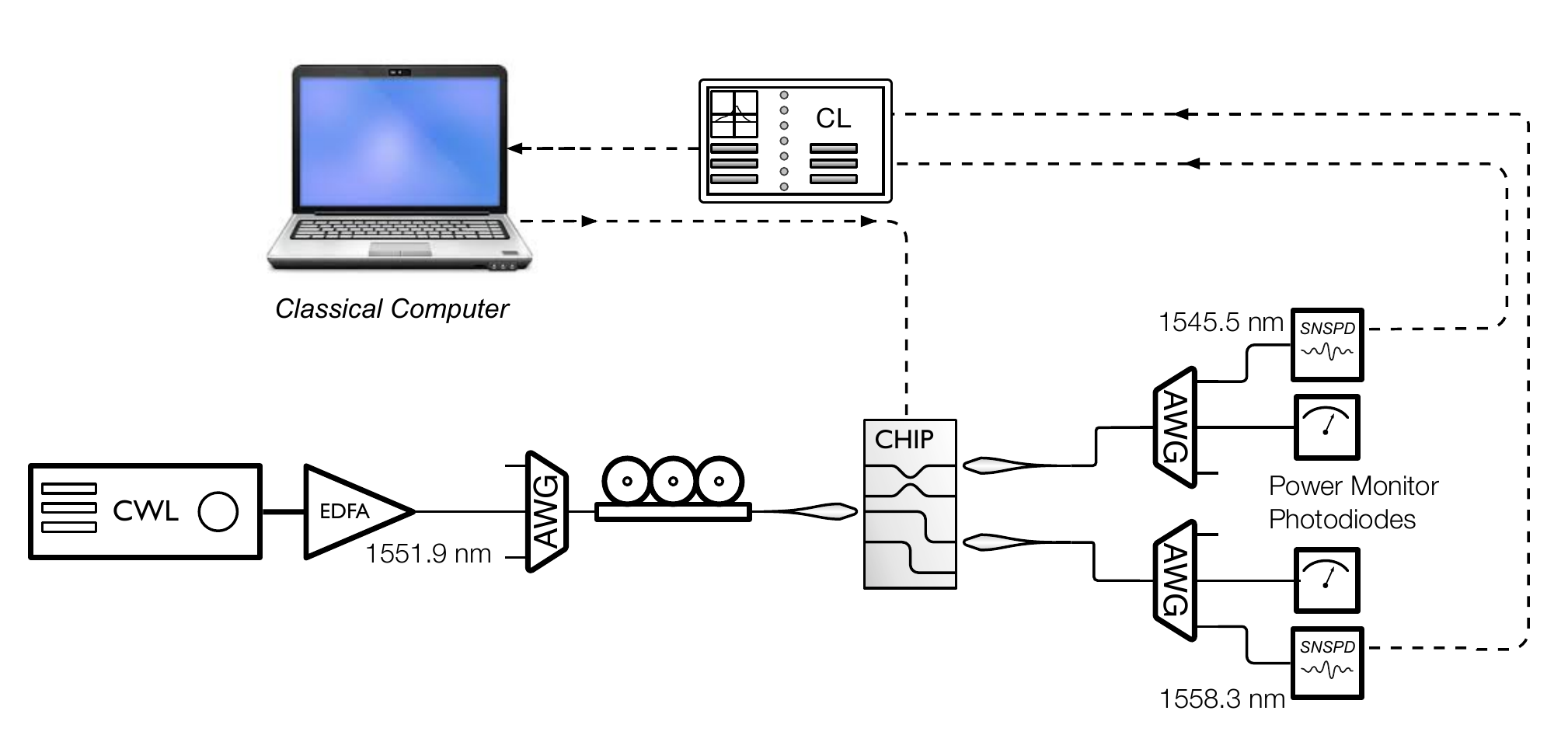}
 \caption{Schematic representation of the experimental set-up. The light from a CW laser light source is amplified through an EDFA. The pump is filtered through an AWG and the polarisation adjusted through a manual polarisation controller, then focused inside the chip and collected by lensed-fibres. The emerging light is then filtered through AWG, where the single photons channel and the the pump are separated. While the CW light is detected by standard photo-diodes for monitoring the coupling, the single photons are detected through super-conductive nano-wire single photon detectors SNSPDs, whose pulses are analysed by a time interval analyser connected to a classical computer. The classical computer is also used to control the on-chip thermo-optic phase shifters.
 }\label{FigSupplExpScheme}
 \end{figure}

 The device is a silicon-on-insulator chip fabricated on a silicon-on-insulator material system using $248~nm$ deep-UV photolithography and dry etching. Light is horizontally coupled into and out of the device via spot-size converters with $300~\mu m$ long inverse taper with a $200~nm$ wide tip beneath a $4 \times 4~\mu m^2$ polyimide waveguide. Approximately $8~dB$ coupling loss per facet was observed between the chip and a single-mode lensed fiber with a $3~\mu m$ mode-diameter. Inside the chip, the photons propagate in integrated single-mode waveguides, designed with a width of $450~nm$ and thickness of $220~nm$ and covered with a $1 ~\mu m$ silicon dioxide upper cladding. Multi-mode interferometers (MMIs) with a footprint of $2.8~\mu m \times 28~\mu m$ were used to realize integrated beam-splitters with near $50\%$ reflectivity.

To stabilise the temperature of the silicon device, the chip was mounted and wire-bonded on a PCB that was glued by thermal hypoxy on a Peltier device connected to a heat-sink. The temperature was controlled by a PID algorithm. 

The input/output optical fibers were automatically recoupled before each scan of the algorithms to maximize the coupling efficiency between waveguides and fibers. 

The device integrates key functionalities for photonic quantum information processing, such as sources to generate an entangled photon-pair and arbitrary single- and two-qubit gates. The photons are generated in two spiral waveguide sources via spontaneous four-wave mixing (SFWM), achieved exploiting the intrinsic $\chi_{(3)}$ non-linearity in silicon~\cite{Silverstone2015,Wang2015}. The two spirally-wrapped waveguide sources were designed with a $1.2\ cm$ length to enhance the photon-pair generation. In the device the qubits are realized encoding the information in the path of the photons~\cite{Politi2008}. The sources were pumped using a continuous-wave (CW) bright laser at $1551.9~nm$, amplified by an erbium doped fiber amplifier (EDFA), with an emerging power of $\sim 10~mW$. Using a fiber-polarization controller before the silicon device we ensured that the transverse-electric (TE) polarised CW light was injected into the device, after the background was removed through the use of array waveguide gratings wavelength division multiplexers (WDM).

Equivalently to the scheme presented in \cite{Wang2015}, the two coherently pumped sources generate a pair of path-entangled photons, i.e. \textit{signal} and \textit{idler} photons, via the SFWM process.  The wavelengths of the signal and idler photons are selected at $1545.5~nm$ and $1558.3~nm$, respectively. Considering the use in this experiment of non resonant SFWM photon pair sources and of a CW laser, the generation probability of multi-photon-pair events is negligible. The output photons, after evolving through the integrated device, were coupled into single-mode fibers, separated apart from the pump light by off-chip AWG with a $0.9~nm$ bandwidth and $> 90~dB$ extinction. Photons were finally detected using superconducting PhotonSpot$^{\text{TM}}$ nano-wires single-photon  detectors (SNSPDs), with system detection efficiency higher than $85\%$, sub-100Hz dark counts, approximately $70$ ps FWHM timing jitter and  recovery times of approximately $50$ ns.
The coincidence counts were recorded by the use of a time interval analyser (Picoharp 300 by PicoQuant$^\text{TM}$) with an integration time of $10~s$ per data point and a coincidence window of approximately $400~ps$. A maximal photon-pair rate of approximately ($200~Hz$) was observed. 

The active and reconfigurable quantum operations inside the silicon device rely on thermo-optical phase shifters, which are formed by metal resistive heaters on top of the silicon waveguides isolated by a dioxide layer in between.  The heat locally dissipated in the waveguide induces the refractive index change responsible of the phase-shift.  The heaters can be independently driven and controlled by home-designed multichannel electronics interfaced with a classical computer. This allows to control the phase with a precision of approximately $0.01~rad$, as reported in more details later on.

\subsection{State evolution and scheme for arbitrary controlled-unitaries}

The logic circuit for the iterative phase estimation algorithms, both RFPE and IPEA, requires a controlled-unitary between the two qubits, as shown in Fig.~1a of the main text. In this section we describe in more details the entanglement-based scheme we used to obtain arbitrary controlled operation in the quantum photonic device.\\
Coherently pumping the two spiral SFWM sources, a pair of photons is generated in a superposition between them.
Using a Fock representation, the photon number state generated is written as
$(|0200\rangle+|0020\rangle)/\sqrt{2}$. Two MMIs beam-splitters following the sources probabilistically split the photons. Post-selecting the cases where the signal photon is collected from the two upper modes and the idler from the four bottom modes, analogously to the scheme shown in \cite{Wang2015, Santagati2016,Zou2013}, the state becomes:
\begin{equation}
\label{FockState}
\frac{|1010\rangle+|0101\rangle}{\sqrt{2}}.
\end{equation}

To encode the qubit in the photons' path, for each photon we use the convention $|0\rangle\leftrightarrow|10\rangle$ and $|1\rangle\leftrightarrow|01\rangle$ to convert the number state to the logic state. Using this convention, the state obtained from the sources can be rewritten as the maximally-entangled Bell state $(|0\rangle_C|0\rangle_P+|1\rangle_C|1\rangle_P)/\sqrt{2}$.\\
Inserting an additional degree of freedom to the path of the idler photon, we obtain the entangled state $(|0\rangle_C|0\rangle_T|0\rangle_P+|1\rangle_C|1\rangle_T|1\rangle_P)/\sqrt{2}$. The two  components associated to the second qubit go through one $R_z$ and one $R_y$ gate to prepare the qubit in the same state $|\psi\rangle_T$ for both pairs of paths. These two components then go through two different operations: $\hat{I}$ for the upper pair of modes in the target register and $\hat{U}$ for the bottom ones, obtaining an entangled state of the form:
\begin{equation}
\frac{ |0\rangle_C|\psi\rangle_T|0\rangle_P+|1\rangle_C(\hat{U}|\psi\rangle_T)|1\rangle_P}{\sqrt{2}}.
\end{equation}
To obtain a superposition of these two different operations, we erase the path information between the two components of the target state.
This is obtained employing two waveguide crossings and combining the modes in two MMI beam-splitters, giving
\begin{equation}
\frac{ (|0\rangle_C|\psi\rangle_T+|1\rangle_C\hat{U}|\psi\rangle_T)|0\rangle_P+(|0\rangle_C|\psi\rangle_T-|1\rangle_C\hat{U}|\psi\rangle_T)|1\rangle_P} {2} .
\end{equation}
Finally, projecting the third qubit into $|0\rangle_P$, the final state can be represented as
\begin{equation}
\frac{ |0\rangle_C|\psi\rangle_T+|1\rangle_C\hat{U}|\psi\rangle_T } {\sqrt{2}}
\end{equation}
which is equivalent to apply the desired arbitrary control-unitary operation.

We remark that this is a post-selected scheme, which makes it not scalable. However, it enables a wide range of experiment previously inaccessible to integrated quantum photonics.

\section{Estimating and simulating phase errors in the integrated 
photonic device}
\label{SimErrors}

\begin{figure}
\centering
\includegraphics[width=1\linewidth]{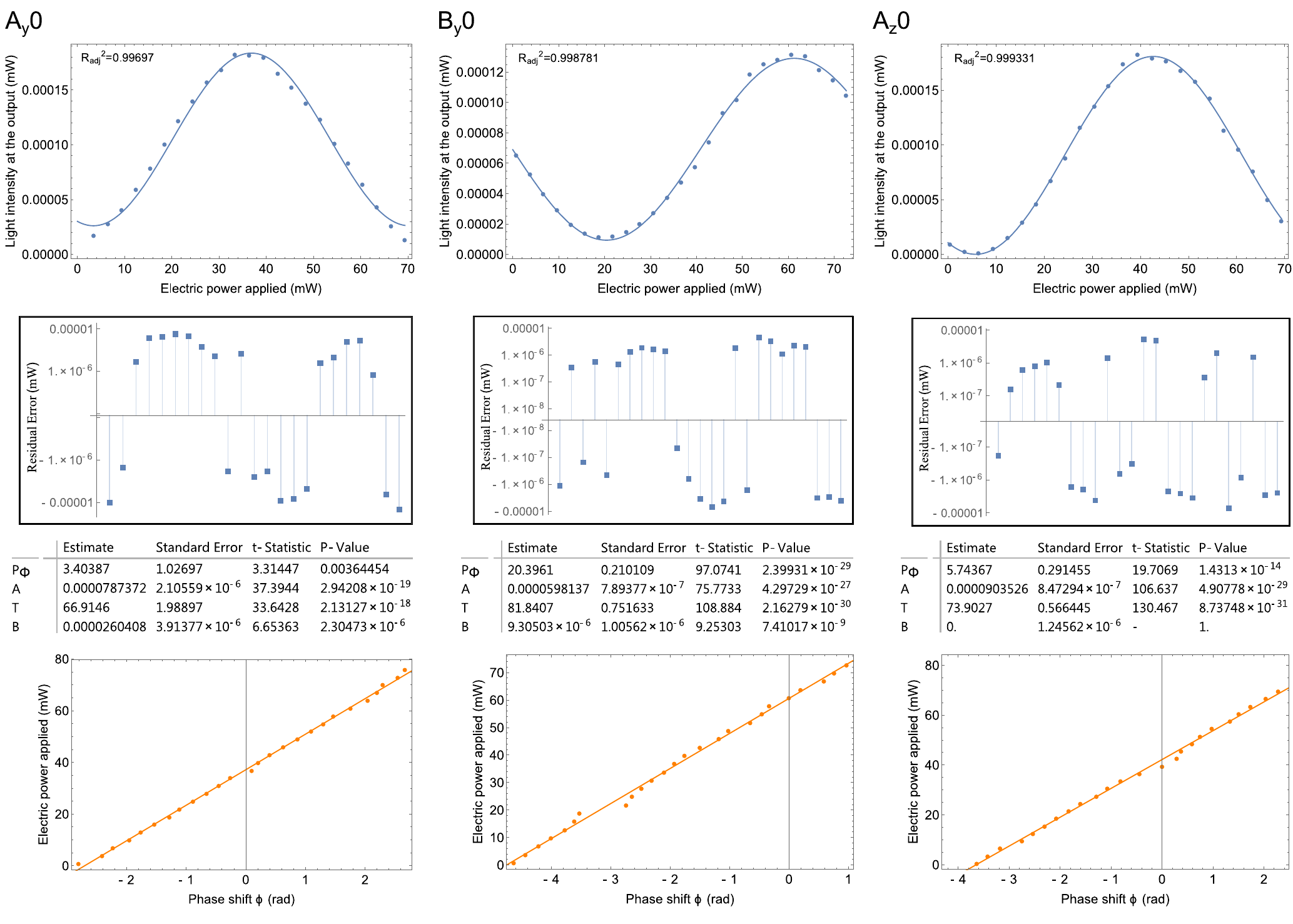}
\caption{Characterization for the implemented phase on three exemplary phase shiifters on the device, chosen as a worst case ($A_y 0$), an average case ($B_y 0$) and a best case ($A_z 0$). For each phase shifter, we report in column, from top to bottom: i) the nonlinear fit of raw experimental data, collected applying electrical power to the phase shifter, while measuring the optical power outputted at the corresponding optical port (vertical axis). The fitting is performed using the nonlinear Eq. \ref{eq:phasecalib0}. The value of $R^2$ is reported in the plot for each separate case. ii) A plot of the residuals corresponding to the nonlinear fits mentioned before. iii) A table of the parameters obtained in the same nonlinear fit is displayed. For each free parameter in the nonlinear model, we report an estimate of both the fitted value, as well as its statistical uncertainty (Standard Error), and a test of its significance (t-Statistic and P-value). iv) For clarity, we report also the linear dependence of the phase shift implemented by each heater against the corresponding electrical power applied.  }
\label{fig:calibration}
\end{figure}

The quantum photonic experimental setup used in this experiment adopts thermo-optical phase shifters to implement both the quantum state preparation and unitary evolution of the qubits. The phase shifters are integrated on-chip as metallic thin-film tracks (depicted with golden rods in Fig.~1B), coating the top of the silicon waveguides. When supplied with DC current, the phase shifters act as non-Ohmic resistive heaters, thus changing locally the refractive index of the waveguide~\cite{Reed2010}. 
Each of the heaters is driven and controlled independently using multi-channel current drivers, in order to reduce electrical cross-talk among the heaters. Cross-talk is expected because of a shared ground connection in the quantum photonic circuit design. Each driver was hosted in a board featuring 12 bits DAC, controlled by the experiment software via RS-232 interface.

Thermal cross-talk among the phase shifters may also occur in a thermo-optically controlled device, because of heat flowing across the dioxide layer and the electrical connections to waveguides other than the targeted one. Thermally induced, unwanted changes in the refractive index of other optical paths alter the implemented phase shifts in a non-controllable way, introducing systematic errors in the setup. In order to compensate for this effect, different calibrations for different chip configurations were preliminary run. For each configuration, all the heaters are constantly driven at a certain current, set to implement an appropriate state preparation and/or gate operation, except the heater to be calibrated. This heater is supplied with a range of currents (and thus electrical power, $P_{el}$), as oscillations in optical power $P_{op}$ at the output of the corresponding optical path are recorded. The collected data can be fitted according to the non-linear function
\begin{equation}
P_{op}=B + A \cos [ \frac{2 \pi}{T} (P_{el}- P_\Phi) ]
\label{eq:phasecalib0}
\end{equation} 
where $B$ is a background, $A$ is the maximum amplitude of the signal, $T$ its period, and $ P_\Phi$ is the offset power value for the heater in such a configuration, all obtained from the model fit. 
Detailed calibrations are reported in Suppl.Fig.\ref{fig:calibration} for three examplary phase shifters, where we also report the relevant statistical parameters of each fit. The fits were performed using the built-in functions from \textit{Mathematica}\textsuperscript{ \textregistered}, that estimate the parameters using a (non-linear) least-squares approach, where the optimization is performed using a numerical global optimization method (``RandomSearch'').
All fits show $R^2$ values close to one, thus suggesting that the model can adequately reproduce the data observed from the measurements. Also, we report high t-statistics and low p-values for most of the parameters, testing the significance of all the parameters adopted in the model, with the exception of the aforementioned background $B$ parameter in those cases where the fringe visibility is particularly high (i.e. no significant background is present). In conclusion, the data analysis gives evidence of the suitability of the model in Eq.~\ref{eq:phasecalib0} to describe the physics of our device.

After the calibration, the targeted phase $\bar{\varphi}$ for a heater is obtained driving the heater with $P_{el}$ such that:
\begin{equation}
\bar{\varphi} = \frac{2 \pi}{T} \left( P_{el} - P_\Phi \right)
\label{eq:phasecalib}
\end{equation} 
and therefore, propagation of stochastic errors in $P_{el}, T,  P_\Phi$ affects $\bar{\varphi}$. 
From experimental non linear fits, relative statistical uncertainties are $\sigma_{P_\Phi}/ P_\text{max} \simeq 0.2 \%$ and $\sigma_{T}/T \simeq 1.1 \%$. Therefore, inaccuracies in the current supplied by the driver to each heater ($\pm 0.005 ~mA$), affecting the actual value of $P_{el}$, can be neglected as they are less than $0.04 \%$ for all the heaters, in the standard configuration used for phase estimation experiments in this paper.

\begin{figure}
   \centering
 \includegraphics[width=0.5\textwidth]{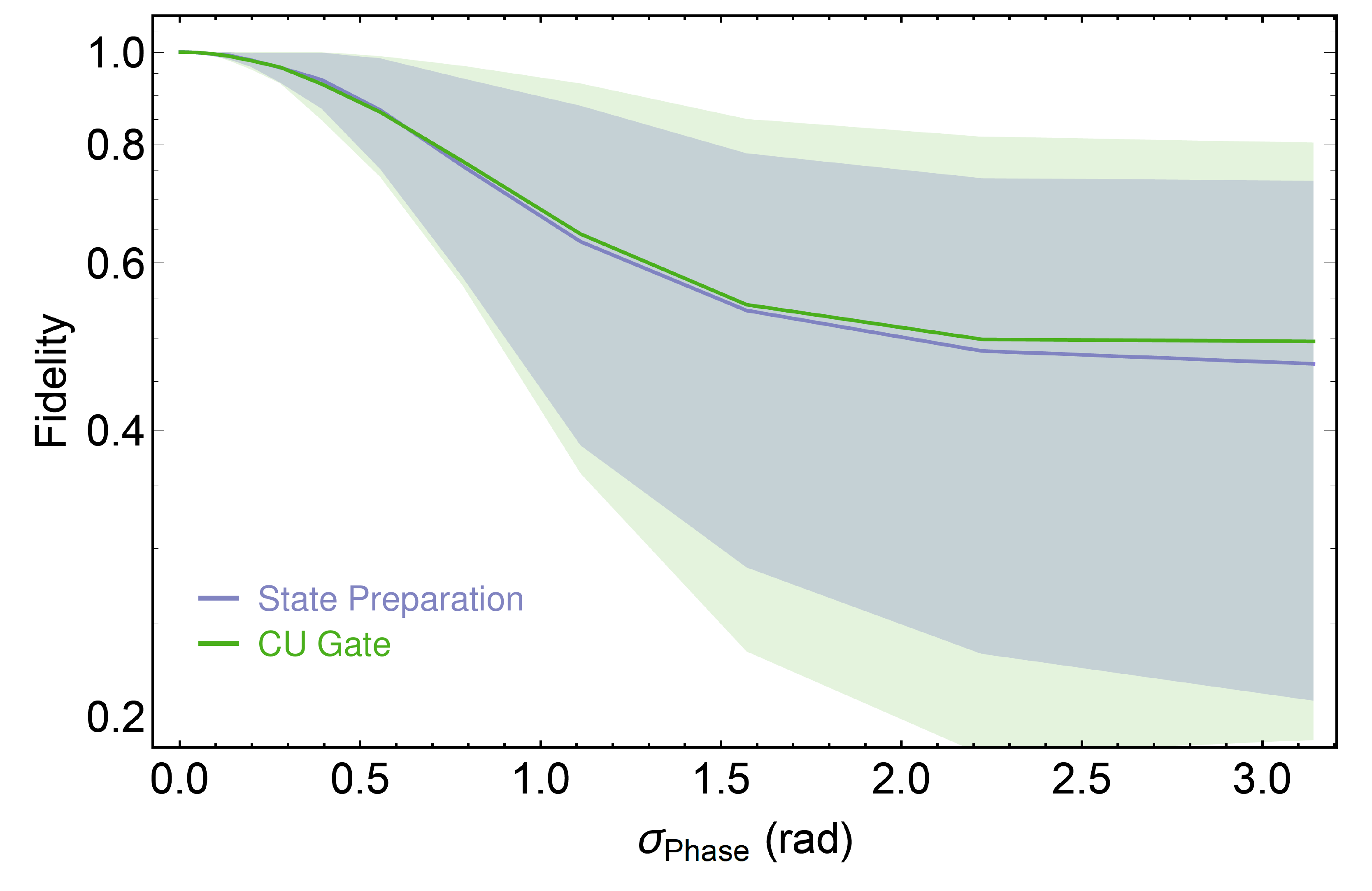}
 \caption{
 	{Simulation of the quantum photonic device performances in terms of the average state preparation fidelity (violet) and $CU$-gate fidelity (green) for different levels of Gaussian noise in the phase shifters, as described by $\sigma_{phase}$ spanning the full interval of values tested in the experiments reported in the main paper. The behaviour for the simulated fidelity of the two operations is very similar, with small discrepancies. }
 }\label{Fig1Supp}
 \end{figure}

Confining the propagation of errors to only $T,  P_\Phi$, and averaging over the full interval of $P_{el}$ adopted in the experiment (appr. $ 5-80 ~mW$, slightly different for each heater due to calibration), an average precision $\sigma_{exp} \simeq 0.01 ~rad$ 
can be estimated for the phases, as experimentally implemented by the heaters. 
In conclusion, systematic errors in the setup can be drastically reduced via an accurate calibration procedure. Nevertheless, in our experiment the effective phase $\varphi$ implemented by each of the phase shifters is affected intrinsically by a stochastic uncertainty $\sigma_{exp}$. This can be estimated from the fits of experimental data. A natural way to investigate the role of imperfect calibration in our device is then to synthetically replace each of the $\bar{\varphi}$, implementing the correct state and unitary preparation, with a phase value sampled from a Gaussian distribution characterized by a variance $\sigma_{phase} > \sigma_{exp}$. An increase in the stochastic noise introduced ($\sigma_{phase}$) corresponds to a decrease in the fidelities for state preparation and gate implementation. A simulation of this dependency is reported in Fig.~\ref{Fig1Supp}

\section{Photonic implementation of RFPE}
\label{MethodsDiscussion}

In the original proposals both IPEA and RFPE algorithms require a single-shot measurement from the experiment~\cite{Kitaev1996,Dobsicek2007,Wiebe2016}. However, in quantum photonics and other quantum platforms, e.g. diamond NV$^-$ centres, measurements usually return a number $N$ of samples distributed according to the output distribution rather than single-shot data, where $N$ depends linearly on the integration time for the data collections. In our experiment photon counts were collected along 10$s$ integration time, leading to typical total counts of $\sim$~2000 for each measure, well beyond the single-shot case. These types of measurements are clearly more informative than the single-shot ones. Nevertheless, phase estimation algorithm (and particularly adaptive implementations) are designed to extract information from single measurements. A challenge in optical implementations is then to convert bulk measures into single-shot data, effectively preserving their enhanced information. 

In photonic experiments the standard way to extract single-shot data from bulk measurements of a qubit is to classically collapse the corresponding binomial distribution adopting a majority voting scheme, thus deciding each single-shot datum according to the most frequent outcome. That is, if the number of photons measured in the state $|0\rangle$ is higher that those in $|1\rangle$ then the output data is $D=0$, otherwise it is $D=1$. This method makes use of some of the information contained in the bulk measurement, enhancing the stability and robustness to experimental noise of the algorithm subtending the measurements. Historically all photonic implementations of IPEA, including ours, have been performed using this scheme~\cite{Lanyon2010,Santagati2016,OMalley2016}.

While the advantages of using a majority voting scheme are clearly maintained for phase estimation protocol, the validity of the approach for RFPE is more subtle. In fact, a majority voting scheme does not produce single-shot data $D\in\{0,1\}$ distributed according to the distribution $\text{Pr}(D|\phi;M,\theta)=\cos^2(\pi M[\phi-\theta] + D\pi/2)$, thus making likely suboptimal the RFPE algorithm, which adopts the likelihood function $\text{Pr}(D|\phi;M,\theta)$.
For this reason, the adoption of a majority voting scheme can be potentially problematic.
When estimating phases from a photonic device, it slows down the inference process because of inefficiencies in the information extracted. When instead the task is to benchmark photonic implementations against experimental platforms based upon single-shot measurements, a majority voting introduces biases.
This is because each run always observes the most likely measurement outcome and feeds it into the learning process, instead of reproducing the measurement statistics expected from random single instances. On one side, this reduces the possibility to retrieve contradictory results from the measurements, speeding up the inference. At the same time, however, it handicaps adaptive algorithms like RFPE, that are optimized for the case when at each step a single measure is sampled from the likelihood function. 

\begin{figure}
   \centering
 \includegraphics[width=0.5\textwidth]{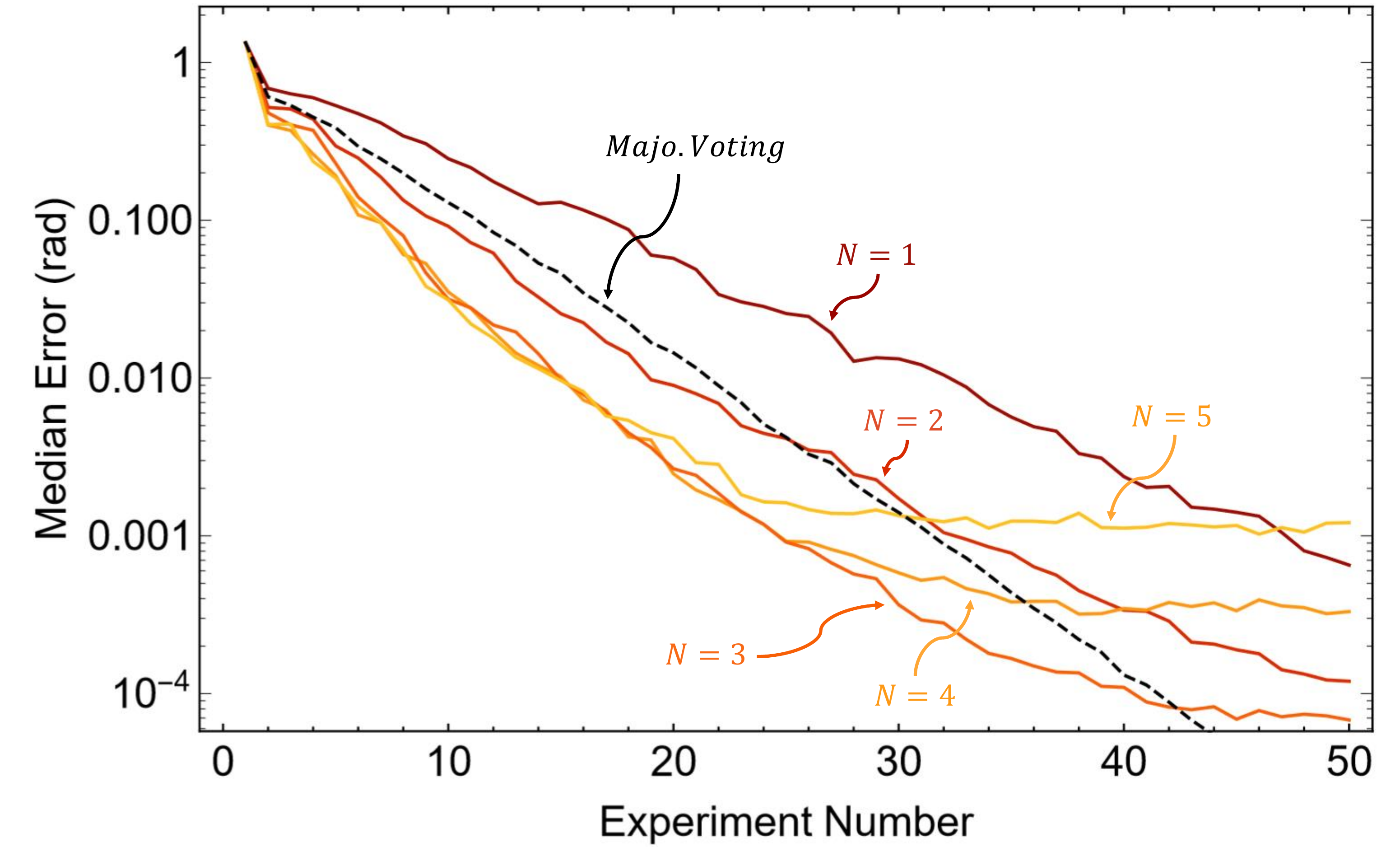}
 \caption{Performance of RFPE using two different strategies: sampling $N$ values from the output distribution measured or using majority voting. The case $N=1$ corresponds to single-shot measurements. Simulated data are averaged over 200 runs.  }\label{Fig2Supp}
 \end{figure}

It is therefore important to compare majority voting with other strategies that could be used to implement RFPE in a photonic environment, whose measurements reproduce better random samples from the true likelihood function, as ideally expected from the algorithm. Results are reported in Fig.~\ref{Fig2Supp}.\\
As an alternative approach, we considered to na{\"i}vely sample $N$ times the value for each single datum from the measured statistics of the photons. This effectively resembles the true likelihood function $\text{Pr}(D|\phi;M,\theta)=\text{Pr}(D|\phi_{\rm true};M,\theta)$. For a high number of photons this is equivalent to sampling the single-shot data from the true distribution of the outputs (in the absence of systematic experimental errors), which is equivalent to the likelihood $P(D|\phi_{\rm true};M,\theta)$. The case where one only photon is sampled randomly from this distribution then coincides with the assumed likelihood function given in the main body for RFPE.  The case where a very large number of photons is sampled corresponds to inferring on the actual data observed, which we process within this framework by using RFPE iteratively on $N$ samples drawn from the empirical likelihood.  Cases that process many sampled photons are not expected to  inherit the asymptotic optimality given by the particle guess heuristic adopted in RFPE~\cite{Ferrie2013,Wiebe2016}. 

We compare the performance of these approaches in  Fig.~\ref{Fig2Supp}.  The first thing to note is that majority voting outperforms the $N=1$ version of RFPE, requiring polynomially fewer measurements to achieve the same fixed accuracy.  This is expected: the inference process in RFPE benefits here from data whose uncertainties are greatly reduced compared to the $N=1$ case.

On the other hand, we observe that small numbers of sampled measurements $N=2,3,4,5$ all outperform majority voting.  
This indicates how already a few random samples give access to a richer information than provided by majority voting at the beginning of the learning process. However, examining later times it is evident how for these same cases RFPE fails to learn in the median for these records.  This diminished robustness is results from taking aggressive experiments in RFPE that lead to posteriors that violate the Gaussian assumptions.  It can be corrected by taking shorter experiments or using the variance reduction strategies in~\cite{Wiebe2016}; however, both constitute adapting RFPE to better process data that naturally arise from photonics experiments

To conclude, these results suggest that though the majority voting approaches commonly used in photonics experiments lead to data records that are atypical of those expected from many quantum computing experiments ($N=1$ case), majority voting provides qualitatively similar results while partially exploiting the enhanced information extraction of photonic measurements.
This, together with the facility to make the comparison with IPEA via the same measurement approach, justifies the use of majority voting for the scope of this work. Finally, considering how all the strategies compared in this paragraph do not fully exploit the entire predictive power of the observed data records and the power of the Bayesian approach implies that there is still space to improve RFPE performances for quantum photonics platform. If we wish to fully assess the power of photonic quantum devices, we need to modify algorithms like RFPE to use likelihood functions that are appropriate for the bulk measurements seen in photonics experiments and transition away from using majority voting as a surrogate for performing proper statistical inference on the data.

\section{Analysis of Breakdown of Majority Voting in IPEA}
\label{IPEAerrors}

\begin{figure}[t!]
\centering
\includegraphics[width=0.5\linewidth]{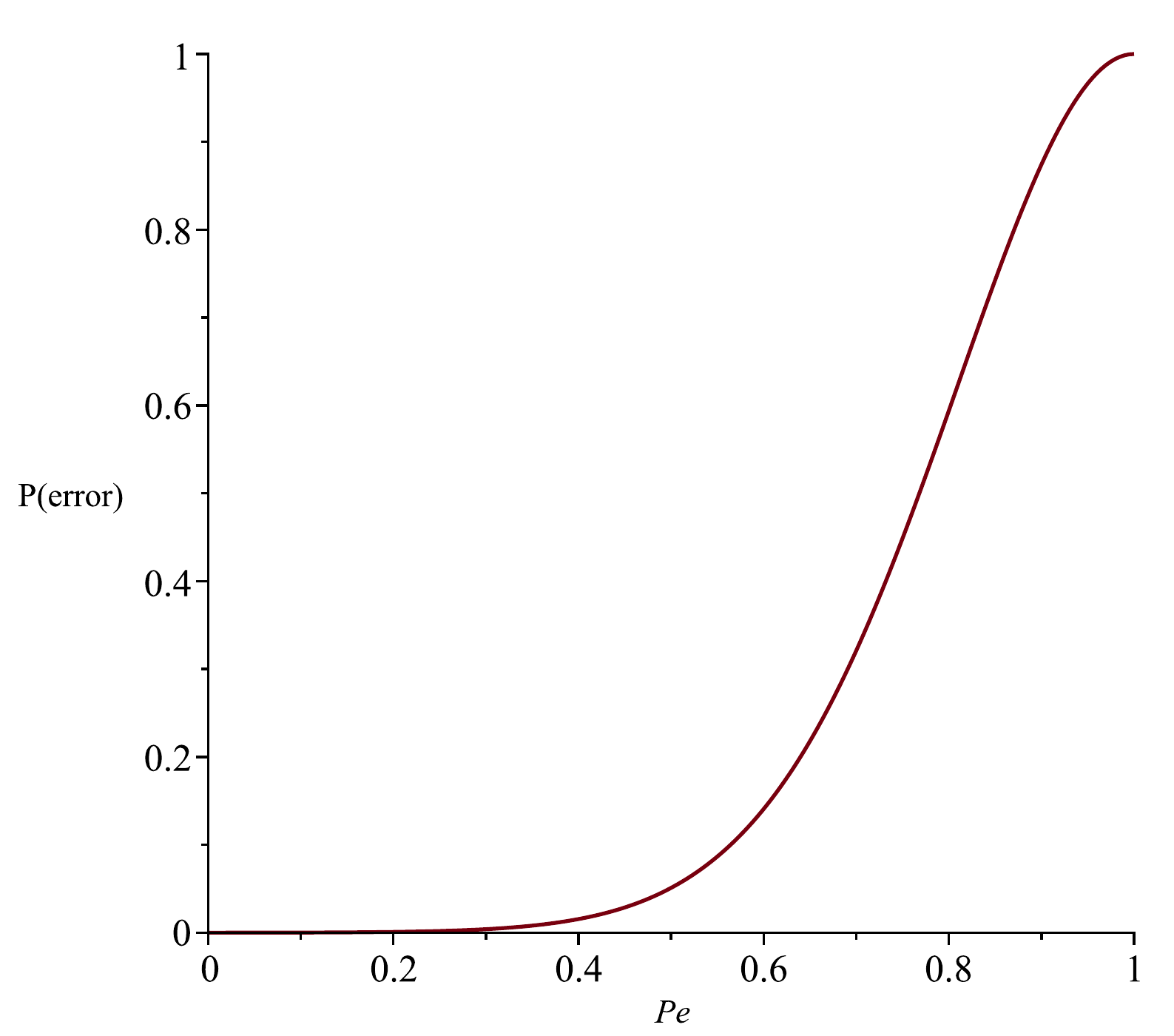}
\caption{Upper bound on error probability of inferring a single bit in a phase estimation experiment given $500$ shots and $P_0=2/3$.  The data shows substantial suppression of errors until $P_e \approx 1/3$ and then rapidly diverges thereafter.  \label{fig:cher}}
\end{figure}

To provide an understanding of why IPEA breaks down so dramatically in the presence of substantial noise, it is instructive to examine a simple model.  Here we consider the model of depolarizing noise, which is similar to that of $T_2$ we considered but for convenience we assume that the strength of the noise is independent of the evolution time.  In particular, let $P$ be the probability of measuring $1$ in an experiment and let us assume without loss of generality that $P\ge 1/2$.  If we define the probability in absentia of noise to be $P_0$ and the probability of an error occurring to be $P_e<1$ then
\begin{equation}
P = P_0(1-P_e) + P_e/2.
\end{equation}

Now let $b$ be the value of a bit inferred from such an experiment and let $n$ be the number of shots used to infer this value by majority voting.  The Chernoff bound then reads
\begin{equation}
P(b=0|P_0>1/2) \le e^{\frac{-n}{2P}(P-1/2)^2}.
\end{equation}
Note that the assumptions that underlie the Chernoff bound are met because $P_0>1/2$ and $P_e <1$ because $P$ is a monotonic function of $P_0$.  As $P_e$ tends to $1$, $P$ approaches $1/2$.  This causes the exponent in the Chernoff bound to tend to $0$.  Thus we no longer gain exponential suppression from $n$ in this limit; whereas, for small $P_e$ the exponent is approximately $\frac{1}{2} \frac{(P_0 -1/2)^2}{P_0} - O(P_e)$.  Thus the exponential suppression of errors expected from majority voting will hold so long as $P_e$ is small and $P_0$ is bounded away from $1/2$.

Now if we assume that the number of bits needed is $n_{\rm bits}$ then the mean number of bits that are incorrect is $n_{\rm bits}e^{\frac{-n}{2P}(P-1/2)^2}$, assuming $P_0>1/2$ for each bit being inferred.
If we solve for the case where the mean number of erroneous bits is at most $1$ we find that
\begin{equation}
P_0 \ge  \frac{1}{2}+ \frac{\sqrt{n_{\rm bits}\ln(n_{\rm bits}) + \ln^2(n_{\rm bits})}-\ln(n_{\rm bits})}{n(P_e-1)},
\end{equation}
suffices to guarantee that this condition is met. 
This shows that the critical signal that we have to have in order to have a reasonable probability of inferring all of the bits rapidly diverges as $P_e$ approaches $1$.  Furthermore, as the number of bits inferred increases the number of shots needed for each inferred bit must increase; although this increase is modest with respect to the number of bits required.

To visualize how this affect plays out, we consider a plot of the probability of erring in the inference of a bit using a sample of $500$ shots (which is easily achievable in our apparatus).  We further assume a value of $P_0 = 2/3$, which again is roughly typical of phase estimation experiments.  We observe that this simple analysis from the Chernoff bound qualitatively reproduces the curves observed in the main body for the error as a function of $T_2$ or the heater variance.  This suggests that the rapid breakdown of IPEA can be understood as a simple consequence of majority voting and the inability of IPEA to detect inference errors and adapt after such an error occurs.

\section{Rescaling the decoherence time $T_2$}
\label{DecTime}

In the decoherence model used $T_2$ represents the decoherence time in units of time required to perform each of the $M$ controlled gates that would be needed to obtain the controlled-$U^M$ operation in a scalable manner. Its value in standard units thus depends on the particular architecture exploited. For example, in a superconductive qubits architecture the operational time for a three qubit controlled-$U$ (using a single Trotter step) is typically $\simeq 1.5~ \mu s$~\cite{OMalley2016}. From our study, this implies that a fully scalable 16-bits IPEA implementation starts to be impractical in superconductive devices with decoherence time approximately lower than $50~\mu s$. Given that the relaxation time of state-of-the-art superconductive qubits is on the same order of magnitude, this highlights how the implementation of IPEA on current pre-fault tolerant devices is very limited.
An equivalent implementation of IPEA on a solid-state electron-spin system would require approximately 50 $\mu s$ for a single controlled-$U$~\cite{veldhorst2015}, and thus 1.5 $ms$ total time for a fully scalable 16-bits IPEA. For such systems, typically observed decoherence times $T_2$ are of the order of tens of $ms$.

\end{document}